\documentstyle[prb,aps,floats,epsfig,preprint]{revtex}

\begin{document}

\title{
  On Pseudogaps in One-Dimensional Models with Quasi-Long-Ranged-Order
}
\author{
  A. J. Millis$^{(1)}$ and H. Monien$^{(2)}$\cite{Bonn}
}

\address{
  $^{(1)}$ Department of Physics and Astronomy\\
  Rutgers University \\
  136 Frelinghuysen Rd\\
  Piscataway, NJ 08854\\
  $^{(2)}$ Theoretische Physik\\
  ETH Z\"urich\\
  CH-8093 H\"onggerberg\\
  Switzerland
}

\maketitle

\begin{abstract}
  We use analytic and numerical methods to determine the density of
  states of a one-dimensional electron gas coupled to a spatially
  random quasi-static back-scattering potential of long correlation
  length. Our results provide insight into the 'pseudogap' phenomenon
  occurring in underdoped high-$T_{c}$ superconductors,
  quasi-one-dimensional organic conductors and liquid metals. They
  demonstrate the important role played by amplitude fluctuations of
  the backscattering potential and by fluctuations in gradients of
  the potential, and confirm the importance of the self-consistency
  which is a key feature of the 'FLEX'-type approximations for the
  electron Green's function. Our results allow an assessment of
  the merits of different approximations: a previous approximate
  treatment presented by Sadovskii and, we show, justified by a WKB
  approximation gives a reasonably good representation, except for a
  ``central peak'' anomaly, of our numerically computed densities of
  states, whereas a previous approximation introduced by Lee, Rice and
  Anderson is not as accurate.
\end{abstract}

\section{Introduction}

''Pseudogaps'', i.e. a suppression of the low-energy electronic
density of states due to an interaction effect, are of current
interest in the context of high-$T_{c}$ superconductivity
\cite{sbproc} and of low-dimensional organic materials \cite{organic}
and are relevant to the theory of liquid metals \cite{Ziman70}, and to
the issue of spin waves above $T_c$ in ferromagnets\cite{Hertz73}. One
general mechanism for producing pseudogaps involves the presence of
long but not infinite range order, for example of the superconducting
or density wave type.  Consider, for definiteness, a one dimensional
material exhibiting long ranged charge density wave order at $T=0$.
At $T>0$ thermal fluctuations prevent long ranged order but at low $T$
the order parameter fluctuates very slowly in space and time.  It
therefore has significant amplitude to back-scatter electrons. The
back-scattering will tend to open a gap but will not do so completely
because the order is not perfect. Very similar issues come up in the
physics of liquid metals \cite {Ziman70}, where the ions are highly
correlated and move slowly in comparison to the electrons, in the
theory of superconductivity, where slow Cooper pairing fluctuations mix
particle and hole states (instead of left and right movers) and again
tend to open a gap, and in ferromagnets at temperatures slightly above
$T_c$, slow magnetization fluctuations may suppress electron-hole spin 
excitations and allow weakly damped spin waves to exist\cite{Hertz73}.

A crucial, but still unresolved, question concerns the proper method
of calculation of physical quantities in the presence of
long-range-correlated scattering. A simple and physically appealing
approximation was proposed in 1973 by Lee, Rice and Anderson \cite
{Lee73}, who argued that one need only consider the leading
perturbative one-boson-exchange diagram for the self-energy.  Within
this approximation the low energy density of states was constant, but
was suppressed from the non-interacting value by a factor proportional
to the inverse of the correlation length of the scatterers. Subsequent
workers used different approximations and obtained different results.
In particular, Sadovskii obtained a continued-fraction expression
which led, for the problem studied in \cite{Lee73} to a low energy
density of states which varied as the square root of the correlation
length \cite {Sadovskii74}. However, Sadovskii's results have recently
been called in to question by Tchernyshyov \cite{Tchernyshyov99} who
exhibited a class of diagrams neglected by Sadovskii. A yet different
set of approximations, the 'fluctuation exchange' or 'FLEX'
approximation \cite{Bickers90} has been employed by many authors to
study pseudogap effects in models of superconductivity \cite{Flex},
and these treatments have in turn been questioned
\cite{Levin98,Tremblay97}.

In order to clarify this situation, we present results of a thorough
numerical and analytical study of a simple model exhibiting pseudogap
effects, namely electrons moving in one spatial dimension and coupled
to a backscattering potential which is constant in time but slowly
varying in space, with correlation length $\xi$. Our main results are
that (1) the approximate treatment of Sadovskii is a good 
approximation to the numerically calculated density of states in the
``high energy'' regime $E > 1/\xi^x$, with $x$ an exponent which
depends on whether the potential is commensurate or incommensurate.
The terms omitted in Sadovskii's derivation noted in
\cite{Tchernyshyov99} evidently produce negligible corrections.  (2)
Neither the 'FLEX' approximation nor non self-consistent treatments
such as that of Lee, Rice and Anderson\cite{Lee73} reproduce at all
well the dependence of density of states on correlation length,
although the self-consistency which is an essential feature of the
FLEX approximation leads to a better representation of the density of
states than does the approximation of Lee, Rice and Anderson.  (3)
Fluctuations in the {\bf amplitude} of the scattering potential play a
crucial role in the form of the low energy density of states. Most of
the models studied make assumptions about the amplitude fluctuations
which are not physically reasonable. (4) Some features of the results
are controlled by fluctuations in the gradient of the backscattering 
potential; the widely used Lorentzian form has divergent gradient
fluctuations which change the form of the results.

The balance of this paper is organized as follows. In section II we
present the models and a simple 'WKB' treatment which reveals the
essential physics.  In section III we present the results of our
numerical study. In section IV we compare the numerical results to
those obtained by approximate methods. Section V contains a summary,
conclusions, and a list of open problems.

\section{Model}

We consider spinless electrons moving in one spatial dimension and
coupled to a static potential with spatial correlations to be
specified below.  The fundamental Hamiltonian is
\begin{equation}
H=-\sum_{j} \left\{ 2 \left(
  d^{\dagger}_jd^{\phantom{\dagger}}_{j+1} + h.c. \right) 
  + \mu d^{\dagger}_j d^{\phantom{\dagger}}_j 
  + V_j d^{\dagger}_j d^{\phantom{\dagger}}_j
  \right\}
\label{eq:H}
\end{equation}

We consider two sub-cases:

A) {\it Half filling and commensurate potential}.  In this case
$\mu=0$ and we assume $V(j)$ is a real random potential chosen from a
distribution which implies
\begin{equation}
< V_j > = 0
\end{equation}
and
\begin{equation}
<V_jV_{j+k}>=(-1)^kf(k/\xi)
\label{eq:Vcommensurate}
\end{equation}

Here $f(x) \rightarrow 0$ for $x \rightarrow \infty$ and we are
interested in the large $\xi$ limit.

B) {\it Incommensurate band filling and potential}.  In this case we
linearize Eq. \ref{eq:H} about the Fermi points, separate the electrons
into right-moving $(R)$ and left-moving $(L)$ branches in the usual
way, and adopt a matrix notation, defining the 2 x 2 matrix ${\bf G}$
via
\begin{eqnarray}
G_{R,R}(z,t;z^{\prime },t^{\prime }) &=&e^{
ip_{F}(z-z^{\prime })}\left\langle Tc_{R}^{\phantom\dagger}(z,t)c_{R}^{\dagger}
(z^{\prime }t^{\prime })\right\rangle  \label{eq:2.1a} \\
G_{L,L}(z,t;z^{\prime },t^{\prime }) &=&e^{-
ip_{F}(z-z^{\prime })}\left\langle Tc_{L}^{\phantom\dagger}(z,t)c_{L}^{\dagger}
(z^{\prime }t^{\prime })\right\rangle  \label{eq:2.1b} \\
G_{RL }(z,t;z^{\prime },t^{\prime }) &=&e^{
ip_{F}(z+z^{\prime })}\left\langle Tc_{R}^{\phantom\dagger}(z,t)c_{L}^{\dagger}
(z^{\prime }t^{\prime })\right\rangle  \label{eq:2.1c} \\
G_{LR }(z,t;z^{\prime },t^{\prime }) &=&e^{-
ip_{F}(z+z^{\prime })}\left\langle Tc_{L}^{\dagger}(z,t)c_{R}^{\phantom{\phantom\dagger}}
(z^{\prime }t^{\prime })\right\rangle  \label{eq:2.1d}
\end{eqnarray}
The Green function obeys a Schroedinger equation which after Fourier
transformation in time may be written: 
\begin{equation}
\left[ \omega +i{\mathbf \sigma}_{3}v_{F}\partial _{z}+{\mathbf V}(z)\right] %
{\mathbf G}=\delta (z-z^{\prime })  \label{eq:2.2}
\end{equation}

A general potential ${\mathbf V}(z)$ has a forward scattering part which
couples right-movers to right-movers and left-movers to left-movers,
and a back-scattering part which couples left movers to right movers.
The forward scattering part is not important for the density of
states\cite{Abrikosov76}; thus we consider only the back-scattering
part and write
\begin{equation}
{\mathbf V}(z)=v_{1}(z){\mathbf \sigma}_{1}+v_{2}(z){\mathbf \sigma}_{2}
\label{eq:2.3}
\end{equation}
Because of the phase factors arising from the incommensurate Fermi
surface $\mathbf{V}$ is a complex potential; further we have absorbed
phase factors into the definition of ${\mathbf G}$, thus we write the $V$
correlator as
\begin{equation}
<V^*_{j}V_{j+k}>=f(k/\xi)
\label{eq:Vincomm}
\end{equation}

Here ${\mathbf \sigma}_{1}=\left( 
\begin{array}{rr}
0 & 1 \\ 
1 & 0
\end{array}
\right) $, ${\mathbf \sigma}_{2}=\left( 
\begin{array}{rr}
0 & -i \\ 
i & 0
\end{array}
\right) $ and ${\mathbf \sigma}_{3}=\left( 
\begin{array}{rr}
1 & 0 \\ 
0 & -1
\end{array}
\right) $ are the usual Pauli matrices.

A formal solution for the Green function of Eq. \ref{eq:2.2} has been
derived by Abrikosov and Ryzhkin \cite{Abrikosov76}; their solution
generalizes immediately to the commensurate case.  The solution is
expressed in terms of the $S-$matrix defined for $z>z^{\prime }$ by
\begin{equation}
{\mathbf S}(z,z^{\prime}; \omega)=i \frac{{\mathbf \sigma}_3}{v_F}%
T_{z}\exp \left( -\int\limits_{z^{\prime }}^{z}dy%
{\mathbf A}(y)\right)  \label{eq:2.3a}
\end{equation}
Here $T_{z}$ is the ''space ordering'' symbol and 
\begin{equation}
{\mathbf A}=\frac{{\mathbf \sigma}_{3}}{v_{F}}\left[ \omega +\vec{v}\cdot \vec{%
{\mathbf \sigma}}\right]  \label{eq:2.4}
\end{equation}
Abrikosov and Ryzhkin show that e.g. the right-moving Green function,
$G_{RR}$, is given for $z>z^{\prime }$ by (we usually do not make the
$\omega$-dependence explicit)

\begin{eqnarray}
\omega >0:\hspace*{0.5cm}G_{RR}(z,z^{\prime }) &=&S_{22}(\infty
,z)S_{22}(z^{\prime },-\infty )/iS_{22}(\infty ,-\infty )  \label{eq:2.5a} \\
\omega <0:\hspace*{0.5cm}G_{RR}(z,z^{\prime }) &=&S_{12}(\infty
,x)S_{21}(z^{\prime },-\infty )/iS_{11}(\infty ,-\infty )  \label{eq:2.5b}
\end{eqnarray}

Further, in this model the density of states for a system of length
$L$ is given by
\begin{equation}
N(\omega)=\frac {1} {L\pi} Tr Im ln {\bf S}(L/2, -L/2; \omega + i \epsilon)
\label{eq:dos}
\end{equation}

Consider the matrix ${\mathbf A}(z)$. Its eigenvalues are 
\begin{equation}
\pm \kappa (z)=\pm \frac{1}{v_{F}}\sqrt{|v(z)|^{2}-\omega ^{2}}
\label{eq:2.6}
\end{equation}
and so we have 
\begin{equation}
{\mathbf A}(z)=\kappa (z){\mathbf Q}(z){\mathbf \sigma}_{3}{\mathbf Q}^{-1}(z)
\label{eq:2.7}
\end{equation}
with ${\mathbf Q}$ a rotation matrix.

If $\kappa^{-1}$ is small compared to the length $\xi$ over which
${\mathbf A}$ varies, ${\mathbf S}$ may be found via a {\em WKB}
approximation.  Details are given in the Appendix. The essential result
is that ${\mathbf S} = {\mathbf Q R D}$ where ${\mathbf Q}(z)$ is the
rotation matrix which diagonalizes ${\mathbf A}(z)$, ${\mathbf R}$ is
a rotation matrix which is close to unity and ${\mathbf D}$ is a
diagonal matrix which, up to a phase, is
\begin{equation}
{\mathbf D} = 
\exp\left( - {\mathbf \sigma}_3 \int\limits^z_{z^\prime} dy \left[
\kappa(y) - \frac{1}{2\kappa(y)} \left[ \left( \partial_y \theta \right)^2 + 
\frac{1}{4} \sin^2(2 \theta) \left( \partial_y \varphi \right)^2
\right]\right] + \ldots  \right)  \label{eq:2.8}
\end{equation}
Here $\theta=tan^{-1}(|v| / \omega)$ and $\varphi =
\arctan(|v_1|/|v_2|)$.

From Eq. \ref{eq:dos} it is clear that G if $\kappa $ is purely real
the density of states vanishes; thus obtaining real excitations of
energy $\omega $ requires the existence of regions in which the local
gap is larger than $\omega $. In particular, in models with only phase
fluctuations, the local gap amplitude $|v|$ is constant and for
$|\omega |<|v|$ the density of states vanishes. This result is of
course familiar in the context of $s$-wave superconductivity, where
low-lying states and indeed dissipation come only from vortices and
phase-slip centers, at which the {\bf amplitude} of the
superconducting order parameter vanishes. The methods used by Lee,
Rice and Anderson \cite{Lee73}, by Sadovskii\cite{Sadovskii74}, and by
many subsequent workers assume that $v$ is Gaussian and peaked at
$|v|=0$ and therefore cannot address the rare amplitude fluctuations
which are the relevant physics for realistic systems as already noted
by \cite{Sadovskii74}.

In the limit of infinite correlation length the density of states
follows immediately from the leading order terms; one finds for a
fixed realization of disorder

\begin{equation}
\frac{N(\omega)}{N_0} =Tr Im {\mathbf G}
=\frac{\omega}{\sqrt{|V(z)|^2-\omega^2}}
\label{eq:dos.def}
\end{equation}

The average density of states is then obtained by averaging this expression
over the probability distribution of $V$, leading to the $\xi \rightarrow
\infty$ results previously obtained by Sadovskii \cite{Sadovskii74}.  For
example, for the incommensurate problem the two independently fluctuating
components of the potential lead to

\begin{equation}
\frac{N(\omega)}{N_0} =\int_0^\omega 
\frac{\Delta d\Delta} {\Delta_0^2} e^{-\frac{\Delta^2}%
{2 \Delta_0^2}}\frac{\omega}{\sqrt{\omega^2-\Delta^2}}
\label{eq:dos.incomm.inf}
\end{equation}

For $\omega \ll \Delta_0$, Eq. \ref{eq:dos.incomm.inf} implies
$N(\omega)/N_0 = \frac{\pi \omega^2} {2 \Delta_0^2}$.  The analogous
expression for the commensurate problem leads to a density of states
proportional to $\omega$.

The expressions presented above suggests that the 'infinite
correlation length' results fail when $\langle (\partial_z \theta
)^2/\kappa \rangle \sim \kappa$, i.e. $\frac{(\partial_z
  |v|)^2\omega^2}{[+ \omega^2 - |v|^2 ]^2} \sim \kappa^2$.  If one
considers this estimate as a function of frequency, then one sees that
in the regions of small potential where low lying states occur,
$\kappa \sim \omega$.  Further, the Lorentzian disorder assumed above
has a divergent second moment, so $\langle (\partial_z V )^2 \rangle
\sim \Delta_0^2 /(\xi a)$ where $a$ is an ultraviolet cutoff of order
a lattice constant.  Combining these factors suggests that the WKB
approximation breaks down for $\omega^4 \sim D^2 (\xi a)$.  Matching
this scale to the low energy density of states $\sim \omega^2$ yields
a residual density of states of order $\xi^{-1/2}$.  As we shall see
in the next section, the numerics suggests rather $\xi^{-1/3}$.
Similar considerations for the commensurate case would yield
$N(\omega) \sim \xi^{-1/4}$; as we shall see, the numerical results
roughly agree. Our arguments suggest that the value of the low energy
density of states is controlled by fluctuations in the derivative of
the random potential.  These are ultraviolet divergent for the widely
used Lorentzian form of the potential fluctuations.  Study of random
potentials with finite second moments and hence derivative
fluctuations on the scale of $\xi^{-1}$ would be of interest.

We now consider the density of states at very low energies.  Very low energies
correspond to very long length scales, and at length scales greater than $\xi$
one expects the problem to map on to one with point-like (delta-correlated)
disorder.  Ovchinnikov and Erikhman \cite{PasturBook} have shown that in a
fluctuating gap model very similar to the $\xi \rightarrow 0$ limit of the
commensurate potential potential case the density of states diverges as
$1/\omega$ (with logarithmic corrections so the integral is not divergent) and
therefore we expect such a divergence in the present commensurate potential
case also.  The interesting question is the dependence on $\xi$ of the
coefficient of the divergent term on the correlation length: this is
equivalent to the question of the frequency range over which the divergence
is visible above the background.

We present here a qualitative argument indicating that the correct
scale is $\xi^{-1/2}$.  We note first that the density of states peak
in the fluctuating gap model may be traced back to the
Su-Schrieffer-Heeger\cite{SuSchriefferHeeger} argument that in such
models a change in sign of the back-scattering potential produces a
mid-gap state. The mean distance between sign changes is $\xi^{1/2}$.
Further, mid-gap states decay exponentially on a scale set by the mean
gap so the hybridization between mid-gap states may be neglected and
they may be treated as independent.  We conclude that the number of
such states per unit length is of order $\xi^{-1/2}$, therefore the
divergence (which is integrable) must exist for $\omega < \xi^{-1/2}$.

This argument may be sharpened.  As noted by Bartosch and
Kopietz\cite{BartoschKopietz}, at $\omega=0$ the Green function for the
commensurate case may be explicitly computed.  In the context of the WKB
formula this may be easily seen from Eq. \ref{eq:2.8}: in the commensurate
case the fact that the potential is purely real means $\partial_z \varphi=0$
while $\partial_z \theta \rightarrow 0$ as $\omega \rightarrow 0$, so
corrections to the WKB result vanish.  The number of states at $\omega=0$ may
then be explicitly computed from Eq. \ref{eq:dos}. At $\omega=0$, ${\bf S}$ is
purely real except when the potential crosses zero, at which point an extra
phase $i\pi$ is incurred. Thus in a system of length $L$ the number of states
at $\omega =0$ is given precisely by the number of zero-crossings, which is of
order $L/(a \xi)^{1/2}$. To convert this into an estimate for the density of
states as a function of frequency, an estimate of the number of nearby states
is required.  We argue that this may be obtained by scaling, assuming $\omega
\sim 1/L$; this leads to a divergence $\sim 1/ \omega (\xi a)^{1/2}$.  Of
course, rough estimates such as these will not correctly capture logarithmic
terms.

Bartosch and Kopietz\cite{BartoschKopietz} have analyzed the formal
solution to the commensurate problem in a different way, obtaining an
expression which they interpret at the zero-frequency density of
states per unit length per unit frequency.  Their expression diverges
exponentially in the size of the system and is consistent neither with
the arguments given above nor with the numerics to be presented in the
next section.  In fact their expression closely resembles the
expression for the real part of $G_{RR}$ which follows from our
analysis. We have argued elsewhere\cite{Millis99} that what they have
computed is a wavefunction amplitude, not a density of states, so we
do not consider their results further.

We now briefly discuss the $\omega \rightarrow 0$ density of states for
incommensurate potentials. The arguments presented above suggest that
this will behave differently than in the commensurate case --the fact
that the fluctuating gap has two components means that the zero
crossing argument is not relevant and the probability for the root
mean square gap to vanish is negligible.  We therefore believe that
the incommensurate model has a vanishing density of states precisely
at zero frequency.

\section{Numerical Results}

In this section we present results of a detailed numerical study of
the density of states of the two models.  We proceed by writing the
real-space form of the Hamiltonian as a matrix, choosing a particular
realization of the backscattering potential from the distribution
defined in the previous section, and numerically diagonalizing the
Hamiltonian.  We obtain the density of states by averaging the
eigenvalues over an appropriate energy window (of order (band
width)/50) and then average over many (typically 1000) realizations of
the disorder. We find that the average over realizations of the
disorder converges more rapidly for longer correlation lengths than
for shorter ones, and converges more slowly for the commensurate model
than for the incommensurate one.  

We first consider the commensurate model, Eq. \ref{eq:H}, which may be
diagonalized as it stands.  For the disorder we chose a Gaussian
distribution in which

\begin{equation}
V_j= Re \left[ \sum_q V(q) e^{iqj} \right]
\label{eq:Vcomm}
\end{equation}
and the distribution of $V(q)$ is determined by the kernel
\begin{equation}
K_{comm}(q)=\frac{\Delta_0^2 sinh(1/\xi)}{cosh(1/\xi)-cos(q-\pi)}
\label{eq:Kcomm}
\end{equation}

These choices correspond to periodic boundary conditions for the
fluctuating potential and open boundary conditions for the electrons.

Because $H$ for the commensurate case may be written as a tridiagonal
matrix, Sturm-chain techniques \cite{Stoer} may be used to obtain
the density of states for very large system sizes (up to $L \sim
10^7$); these system sizes are large enough that boundary effects are
entirely negligible; study of the length dependence of results has has
allowed us to verify that even for the smaller systems ($L \sim 10^4
$) accessible to direct diagonalization, boundary effects are
negligible.

Figure 1 shows the density of states for the commensurate model for several
correlation lengths and $\Delta=0.2$. Several features are immediately
evident.  First, as noted by other workers\cite{Lee73}, correlation lengths
larger than $\xi_{\Delta}=v_F/\Delta$ are required in order to obtain an
appreciable pseudogap.  Second, the density of states drops only slowly as
$\xi$ is creased, and is surprisingly large even at $\xi \sim 100$.  Third, at
low energies the density of states is approximately constant except for a
'central peak' which is centered at $\omega=0$ and has a width which
diminishes as the correlation length increases.  We have obtained the residual
(i.e. without central peak) density of states by smoothly extrapolating the
calculated $N(\omega)$ to zero, neglecting the upturn.  The procedure suffers
from ambiguities because our data are relatively coarsely binned in frequency
and there is some uncertainty about the proper functional form for the central
peak.  The results are shown as filled circles in Fig. 2; the dashed line
clearly shows that the residual density of states vanishes as $\xi^{-\eta}$
with $\eta \sim 1/2$.

We now turn to the central peak. It is evident that the width decreases as
$\xi$ increases. Our data are not sufficiently accurate to allow us to
determine the precise scaling of the central peak with $\xi$ and $\omega$, but
we believe they are consistent with the form $N(\omega, \xi)\sim
f(\omega\xi^{1/2})$ with $f(x)$ given by the Ovchinnikov-Erikhman
form\cite{OvchinnikovErikhman}, $f(x) \sim 1 / x \ln^3(x)$ for $x<1$ and $f
\rightarrow 0$ for $x\rightarrow 1$.  In particular, $N(\omega)$ appears to
increase roughly as $1/\omega$ and as shown in Fig. 3, the area under the
central peak scales approximately as $\xi^{-1/2}$, although one sees that the
corrections to this scaling become appreciable for $\xi\le 100$.

The incommensurate model is defined via a continuum equation; a
discretization is therefore necessary.  For computational convenience
we have adopted
\begin{equation}
H=-\sum_j i \left\{ 
\left[
   d^{\dagger}_{j+1,R} d^{\phantom{\dagger}}_{j,R}
  -d^{\dagger}_{j,R} d^{\phantom{\dagger}}_{j+1,R}) - (\leftrightarrow)
\right]
+V_j \left(
   d^{\dagger}_{j,L} d^{\phantom{\dagger}}_{j,R}
 + d^{\dagger}_{j,R} d^{\phantom{\dagger}}_{j,L}
\right)
\right\}
\label{eq:incomm_numeric}
\end{equation}
with complex $V_j$ gaussian distributed with correlator
\begin{equation}
<V^*_j V_{j+k}>=\Delta_0^2 e^{-|k|/\xi}
\label{eq:Kincomm}
\end{equation}

Our results for the density of states are shown in Fig. 4. The more
rapid drop of $N$ with $\omega$, expected from the WKB argument, is
evident. There is no central peak. Indeed there are some indications of
a `central dip' but it is difficult to resolve this question
numerically. We have obtained the residual density of states by
smoothly extrapolating the calculated $N(\omega)$ to zero, neglecting
any possible downturn; the procedure is less ambiguous than in the
commensurate case. The results are shown as filled squares in Fig.
\ref{fig:dos0-cmp}; the scaling with $\xi$ seems to be closer to
$\xi^{-2/3}$ than to the theoretically predicted $\xi^{-1/2}$.

\section{Comparison to Approximate Calculations}

In this section we discuss the relation of the numerical results to
various approximate calculations, in order to obtain insight into the
strengths and weakness of the different approximations. We begin with
the WKB approximation, shown as heavy dashed line in Figs. 1 and 4.
This is seen to be a good approximation to the calculated density of
states for not too low energies and not too short correlation lengths;
essentially the numerical results follows the WKB one until the
density of states drops to the residual level shown in Fig. 2.

We now turn to the continued fraction method of Sadovskii, which for
the incommensurate case is compared to numerics in Fig. 4. The
qualitative correspondence is seen to be good, and to improve for
longer correlation lengths. This suggest that the terms which
Tchernyshyov\cite{Tchernyshyov99} has noted are neglected in
Sadovskii's approach are not quantitatively important and become less
significant as $\xi$ is increased. For infinite correlation length
Sadovskii's results are justified by the WKB arguments of section II.
Closer examination however shows that the low energy large $\xi$
behavior of the density of states is not so well represented as one
can see in Fig. 2, the magnitude differs from the numerical one by
factors of order two and the scaling with $\xi$ is incorrect, being
$\xi^{-1/2}$ instead of the numerically determined exponent ${-2/3}$
to ${-1}$. Similar discrepancies arise in the commensurate case, where
the Sadovskii $N(0)\sim\xi^{-1/3}$ instead of the correct
$\xi^{-1/2}$.

We now turn to the two other widely used approximations with a more
transparent physical content, namely the Lee-Rice-Anderson (LRA) and
``fluctuation-exchange'' (FLEX). The former authors argued that one
should approximate the electron self energy by the leading order
graphs which correspond to the expression
\begin{equation}
  \label{eq:sigma-LRA}
  \Sigma_{LRA} 
  = \int G_0 K 
  = \int\frac{dq}{2\pi} \frac{K(q)}{i\omega-\epsilon_p+Q_q}
  \approx \frac{1}{i\omega+\epsilon_p+i/\xi}
\end{equation}
where $Q=2k_F$, $G_0=(i\omega-\epsilon_p)^{-1}$ is the bare Green's
function and $K(q)$ is defined in Eqs. \ref{eq:Vcomm} and
\ref{eq:Kcomm}.  This approximation leads to a gap structure which
becomes very sharp even for relatively small $\xi$ and to a low energy
density of states which varies as $1/\xi$ for both commensurate and
incommensurate cases. These incorrect results arise because the self
energy is too singular; this feature in turn arises because the bare
electron Green's function is used to describe the intermediate state.

An alternative scheme is the FLEX method. This is complicated in
general but in the case of present interest is equivalent to making
the Lee-Rice-Anderson calculation self-consistent, by using a fully
dressed Green's function to compute the one-loop self energy i.e.
$\Sigma_{FLEX}(k,\omega) = \int G\;K$ with
$G=(i\omega-\epsilon_k-\Sigma_{FLEX}(k\omega))^{-1}$.

We have numerically solved the FLEX equations for $\Delta=0.2$ and
various $\xi$, the results are shown in Fig. \ref{fig:dos-flex}. It is
clear from the results that the dressed Green function leads to a less
singular integral and therefore to a larger low energy density of
states with less $\xi$ dependence.  These are successes; on the other
hand, as with the LRA approach, the difference between commensurate
and incommensurate cases is lost and the central peak is absent. Also
as seen in Fig. \ref{fig:dos-flex} the location of the peak in $N(E)$
has a strong correlation length dependence, inconsistent with the
numerics and with physical intuition. The value of $N(0)$ is found to
be larger than that found numerically but the $\xi$ dependence is
qualitatively reasonable ($\sim \xi^{-1/2}$).

\section{Conclusion}

To summarize, we have used numerical methods and a {\em WKB} analysis
of a formal solution of a Schroedinger equation to obtain an
expression for the Green function of a model of a one-dimensional
charge density wave in its fluctuation regime. We found that low-lying
density of states comes from regions where the amplitude of the $CDW$
gap vanishes and we emphasize that a proper treatment of a physically 
relevant model requires a correct treatment of amplitude fluctuations,
which are typically described by a non-Gaussian probability distribution,
which is difficult to handle either analytically or numerically.

On the qualitative level, we found that pseudogaps require relatively
extreme conditions: a drop in the density of states does not begin to
appear until the correlation length is larger than the basic
coherence length $v_F/\Delta_0$ defined by the electron velocity and
mean field fluctuation amplitude, and the low energy density of states
decreases only slowly as $\xi$ is increased beyond this scale. Physics
we have omitted from our model, including quantum fluctuations of the
pairing potential and the phase space effects characteristic of
dimensions greater than one, only weakens the tendency to form a gap.
The `pseudogap' observed in underdoped high-T$_{\rm c}$
superconductors involves a significant suppression of the low energy
density of states and therefore implies, at least for these materials,
the existence of well established, reasonably long ranged pairing 
fluctuations.

On the technical side, we have shown that the WKB method (which we
suspect can be generalized to higher dimensions) and the Sadovskii
approximation (which probably cannot) provide relatively reasonable
estimates of the density of states; other approximations do rather
poorly, which is unfortunate because they easily generalize well to
dimensions larger than one.

Three extensions of this work would be desirable. One is to calculate
the conductivity; another is to numerically investigate the crossover
between the large-$\xi$ pseudogap behavior and the small-$\xi$
constant density of states behavior; the third is to treat
non-Gaussian amplitude fluctuations.

{\it Acknowledgments} We thank P. B. Littlewood, P. A. Lee and Boris
Shraiman for helpful discussions. Portions of this work were performed
under the auspices of the Correlated Electron Theory Program at Los
Alamos National Laboratories and portions at the Institute for
Theoretical Physics, and portions at the Johns Hopkins University.
H.M. thanks Bell Laboratories and Johns Hopkins University for
hospitality.  Research reported here was also supported by N.A.T.O.
grant CGR 960680, and NSF DMR 9707701.

\newpage 

\appendix*

\section{WKB approximation for S}

For $z > z^\prime, {\mathbf S}$ obeys the equation
\begin{equation}
\left[ \partial_z + {\mathbf A} (z) \right]{\mathbf S} = 0  \label{eq:A-1}
\end{equation}
with 
\begin{equation}
{\mathbf A}(z) = i \omega {\mathbf \sigma}_3 + \Delta(z) \cos \varphi(z) %
{\mathbf \sigma}_1 + \Delta(z) \sin \varphi(z) {\mathbf \sigma}_2  \label{eq:A-2}
\end{equation}
Here ${\mathbf \sigma}_{1, 2, 3}$ are the usual Pauli matrices and $\Delta$
and $\varphi$ are related to the quantities $v_1$ and $v_2$ defined in the
text by $v_1 = \Delta \cos \varphi$; $v_2 = \Delta \sin \varphi$. The
eigenvalues of ${\mathbf A}$ are $\pm \kappa(z)$ with $\kappa^2(z) =
\Delta^2(z) - \omega^2$
If the scale $\xi$ over which ${\mathbf A}$ varies is much larger than $%
\kappa^{-1}$, Eq. \ref{eq:A-1} may be solved via a {\em WKB} approximation.
Write 
\begin{equation}
{\mathbf S} (z , z^\prime ) = {\mathbf Q} (z) {\mathbf R}(z) {\mathbf D}(z) %
{\mathbf I}(z^\prime )  \label{eq:A-4}
\end{equation}
Here ${\mathbf Q} \exp\left( i \theta \vec{n} \cdot \vec{{\mathbf \sigma}}\right)$
with 
\begin{eqnarray}
\tan( z \theta) &=& \Delta / \omega  \label{eq:A-3a} \\
\vec{n} &\equiv& (n_1 , n_2 , n_3 ) = (\sin \varphi, - \cos \varphi , 0)
\label{eq:A-3b}
\end{eqnarray}
and 
\begin{equation}
{\mathbf D} = \exp\left( - \int\limits^z_{z^\prime} dy \left\{ \left[ \kappa(y)
+ d_3 (y) \right] {\mathbf \sigma}_3 + d_0 (y) \right\} \right)  \label{eq:A-5}
\end{equation}
is a diagonal matrix and $d_3$ and $d_0$ are functions to be determined. $%
{\mathbf R}$ is a rotation matrix which is close to the unit matrix and $%
{\mathbf I}$ expresses the initial conditions. Using Eq. \ref{eq:A-4} and Eq.
\ref{eq:A-5} in Eq. \ref{eq:A-1} gives
\begin{equation}
[\kappa + d_3]{\mathbf R \sigma}_3 {\mathbf R}^{-1} + d_0 - (\partial_z %
{\mathbf R} ) {\mathbf R}^{-1} = 
[{\mathbf Q}^{-1} \partial_z {\mathbf Q} ] + \kappa %
{\mathbf \sigma}_3  \label{eq:A-6}
\end{equation}
Explicitly, $[{\mathbf Q}^{-1} \partial_z {\mathbf Q}] = i \vec{q} \cdot \vec{%
{\mathbf \sigma}}$ with 
\begin{equation}
\vec{q} = \vec{n} \partial_z \theta + \frac{1}{2} \sin 2 \theta \left( \vec{n%
} \times \hat{z}\right) \partial_z \varphi + \sin^2 \theta \hat{z} \varphi_z
\label{eq:A-7}
\end{equation}
By assuming ${\mathbf R}$ is the unit matrix ${\mathbf 1}$ plus small
corrections and iterating the equation one obtains
\begin{eqnarray}
d_3 &=& i \sin^2 \theta\;\partial_z \varphi - \frac{1}{2\kappa} \left[
(\partial_z \theta)^2 + \frac{1}{4} \sin^2(2 \theta)(\partial_z
\varphi)^2\right] + \ldots  \label{eq:A-8a} \\
{\mathbf R} &=& {\mathbf 1} + \frac{i (\hat{z} \times \vec{q})}{2 \kappa} 
- \frac{(%
\hat{z} \times \vec{q})^2}{8 \kappa^2} + \frac{i \hat{z} \times \partial_z}{
\kappa} \left[ \frac{\hat{z} \times \vec{q}}{2 \kappa} \right] + \ldots
\label{eq:A-8b}
\end{eqnarray}
This solution may obviously be extended. A non-zero value of $d_0$ occurs in
the third order.

\newpage
\section*{Figure Captions}
\begin{itemize}

\item[Fig. 1] Energy (E) dependence of density of states $N$ calculated
  numerically for commensurate potential model defined by Figs. 1-3 of the
  text, for different correlation lengths $\xi$. The mean gap value $\Delta_0
  = 0.2$ is shown by a dotted line. The data come from eigenvalues binned into
  intervals of width 0.01.  The dashed line shows the `infinite correlation
  length' result from the WKB approximation.
  
\item[Fig. 2] Residual $(E \to 0)$ density of states $N$ calculated
  numerically for the commensurate potential defined by Eqs. 1-3 (filled
  circles), the incommensurate potential defined by Eqs. \ref{eq:Vincomm}
  (diamonds), and the approximation of Sadovskii for commensurate
  (triangle) and incommensurate (square) cases, are plotted against
  correlation length $\xi$ on doubly logarithmic axes. Also shown are dotted,
  dashed, and dash-dotted lines indicating scaling with $\xi^{-1/3}$, $\xi^{-
    1/2}$, $\xi^{- 2/3}$, respectively. For the numerically calculated
  commensurate potential case, the central peak evident in Fig. 1 has been
  subtracted as described in the text.

\item[Fig. 3] Spectral weight, i.e. integrated area under central peak for the
  commensurate potential case, plotted against square root of inverse
  correlation length.
  
\item[Fig. 4] Energy (E) dependence of density of states $N$ calculated
  numerically for incommensurate potential model defined by Eqs. 1, 22, 23 of
  the text,  for different correlation lengths $\xi$. The mean gap value
  $\Delta_0 = 0.2$ is shown by the vertical dotted line. The results are
  obtained from numerically calculated eigenvalues binned into intervals of
  width 0.01. The dashed line is the result of the the `infinite correlation
  length' WKB calculation.
  
\item[Fig. 5] Energy (E) dependence of density of states $N$ calculated via
  the FLEX approximation from Eqs. \ref{eq:sigma-LRA} of the text for different
  correlation lengths and the incommensurate potential. The infinite
  correlation length WKB result is shown by the dashed line. Note that the
  mean gap value $\Delta_0$ was fixed at 0.2; the variation of the position of
  the density of states peak is an artifact of the FLEX approximation.

\end{itemize}

\newpage
\begin{figure}[t]
  \protect\centerline{\epsfig{file=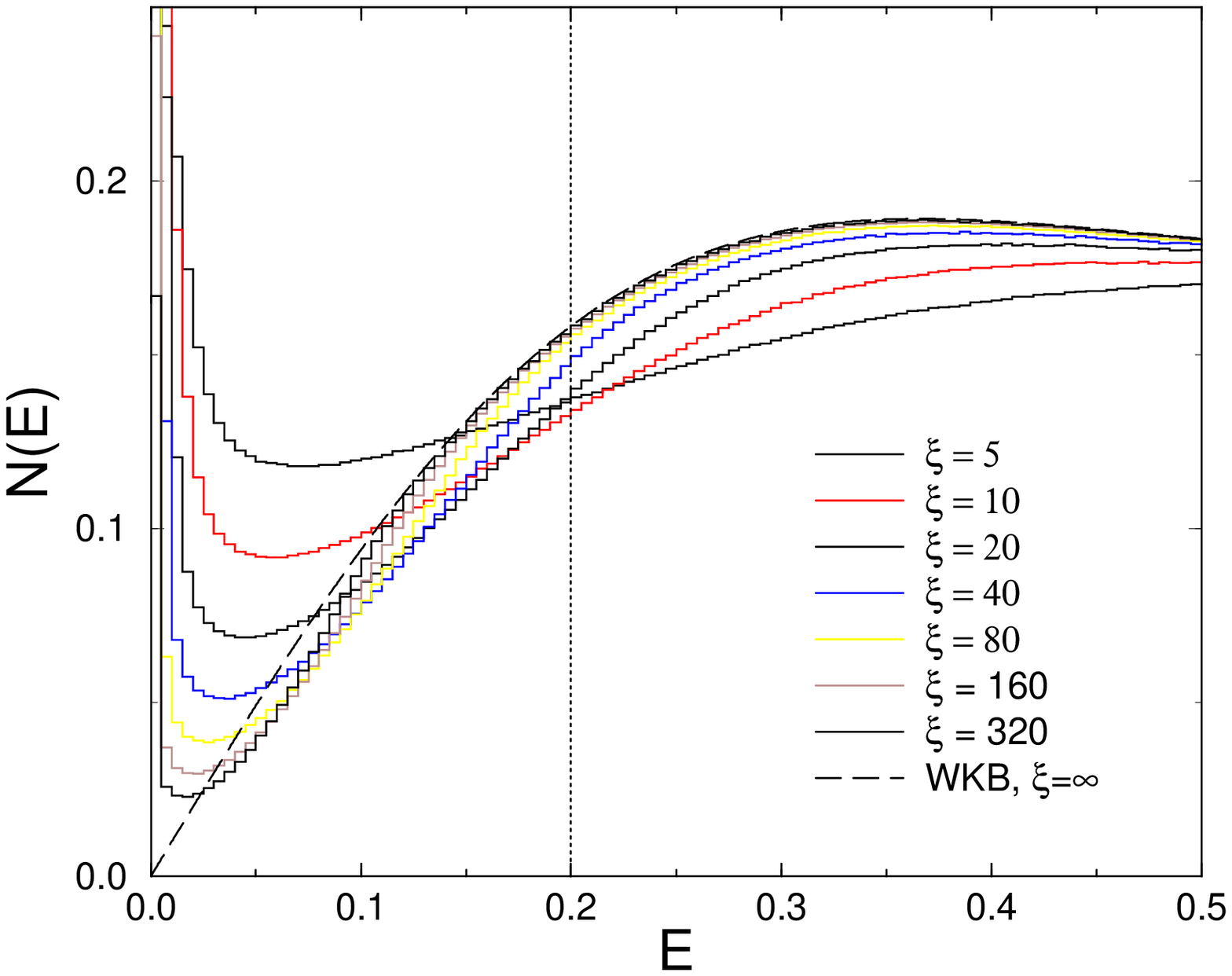,height=14cm}}
  \protect\caption{}
  \label{fig:dos-comm}
\end{figure}

\begin{figure}[t]
  \protect\centerline{\epsfig{file=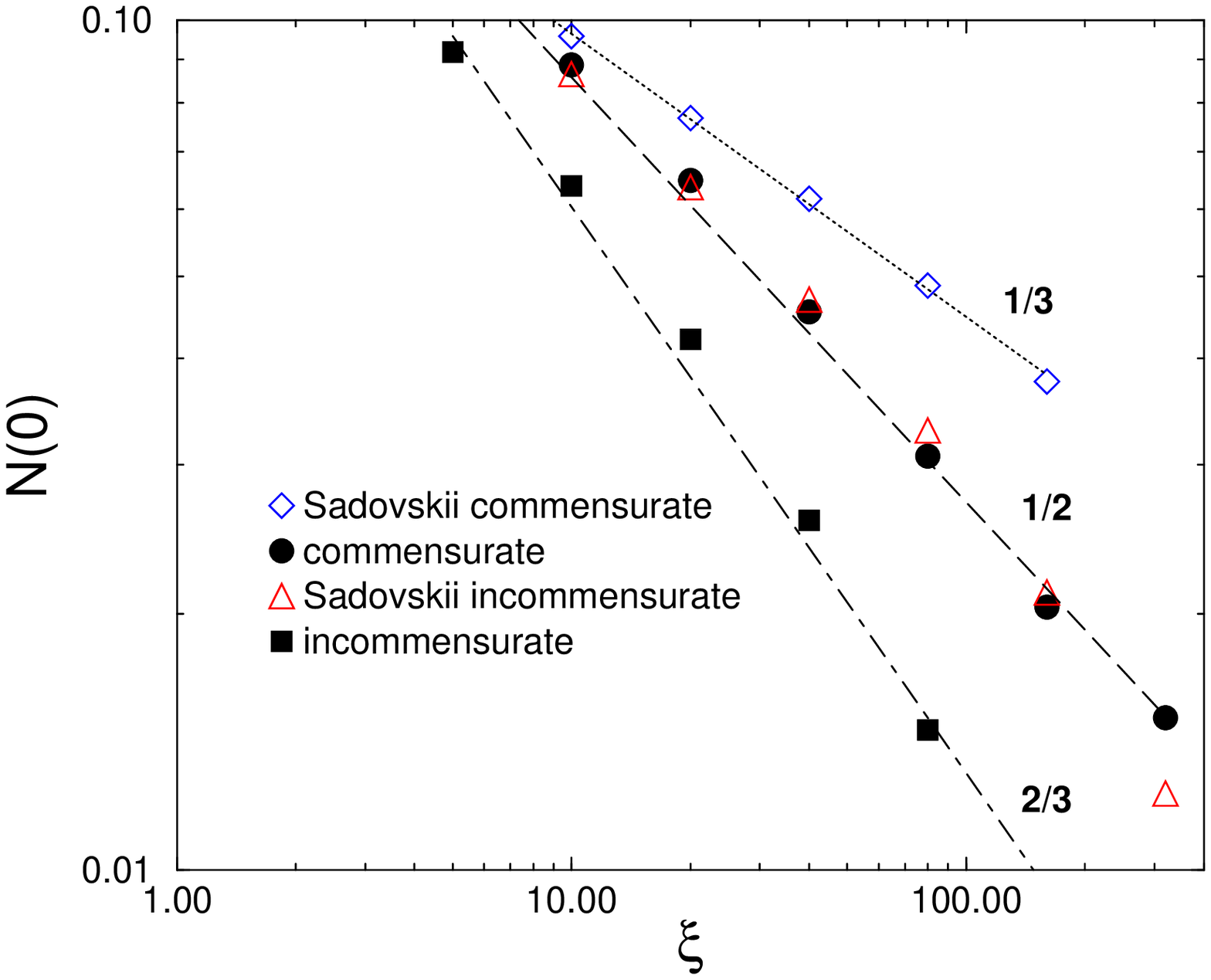,height=14cm}}
  \protect\caption{}
  \label{fig:dos0-cmp}
\end{figure}

\begin{figure}[t]
  \protect\centerline{\epsfig{file=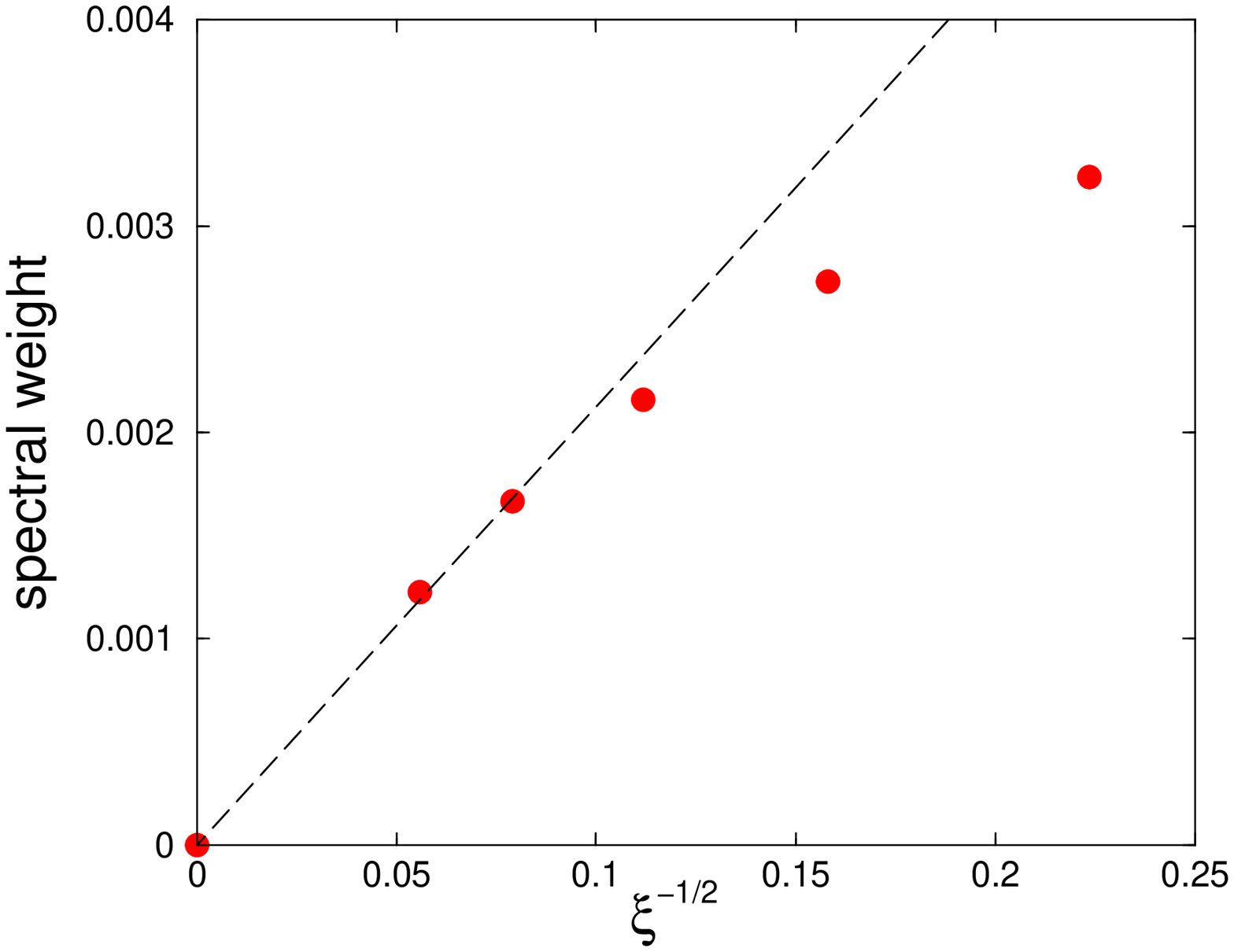,height=14cm}}
  \protect\caption{}
  \label{fig:spectral-weight}
\end{figure}

\begin{figure}[t]
  \protect\centerline{\epsfig{file=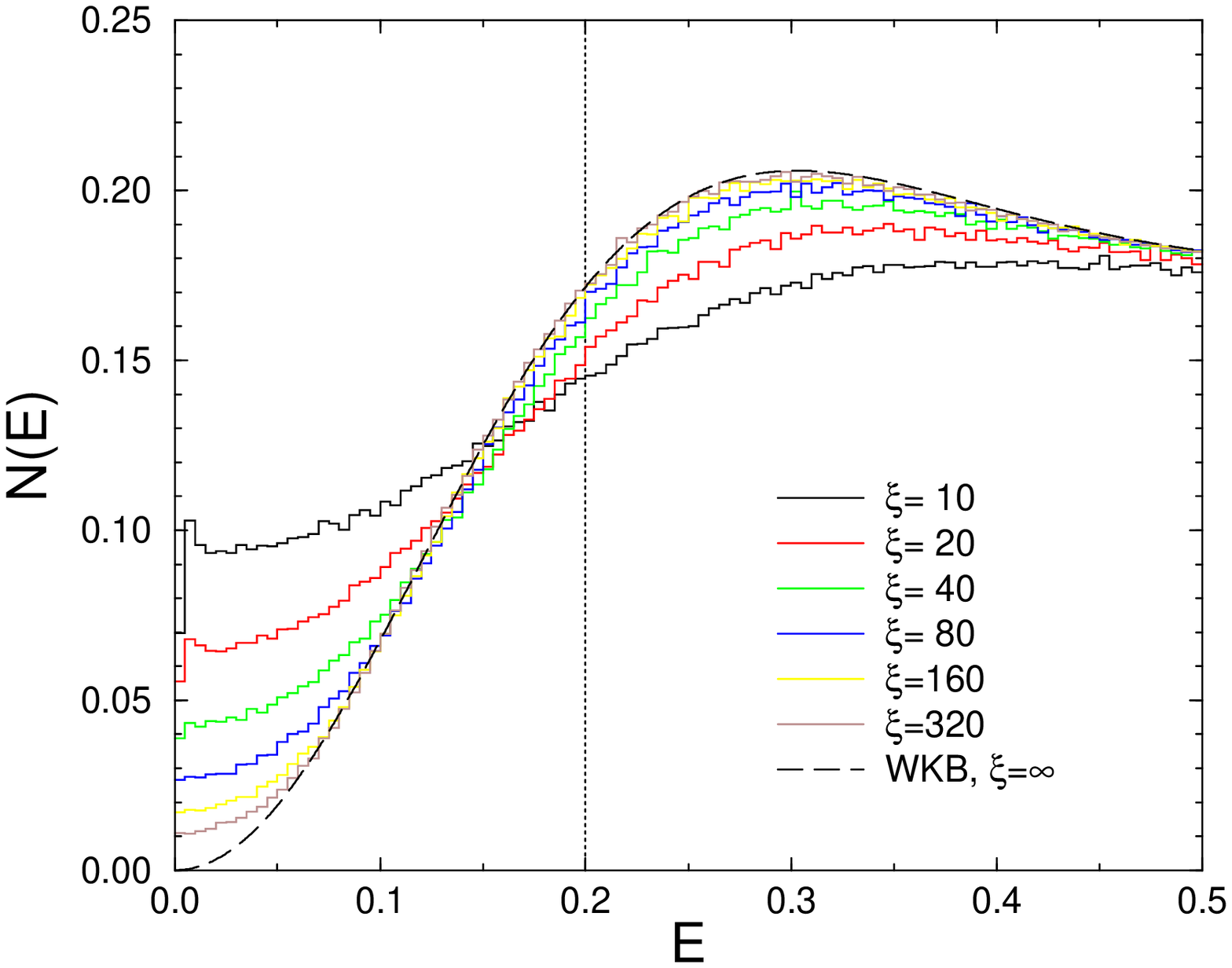,height=14cm}}
  \protect\caption{}
  \label{fig:dos-inc}
\end{figure}

\begin{figure}[t]
  \protect\centerline{\epsfig{file=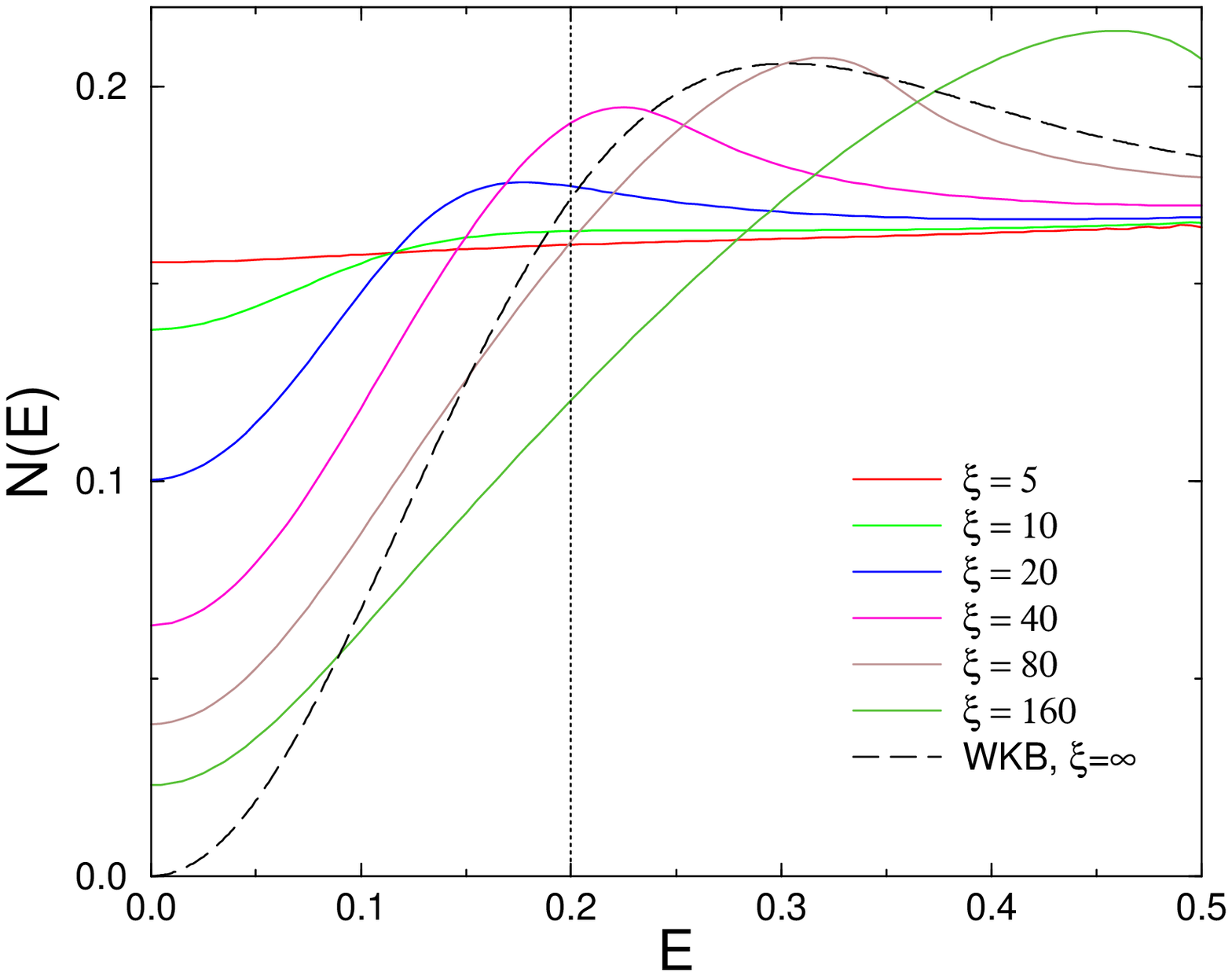,height=14cm}}
  \protect\caption{}
  \label{fig:dos-flex}
\end{figure}

\end{document}